\documentclass[a4paper,fleqn,usenatbib]{mnras}
\usepackage{savesym}
\usepackage{amsmath}
\savesymbol{iint}
\savesymbol{iiint}
\usepackage{txfonts}
\restoresymbol{TXF}{iint}
\restoresymbol{TXF}{iiint}

\usepackage[T1]{fontenc}
\usepackage{ae,aecompl}

\usepackage{graphicx}
\usepackage{amsmath}
\usepackage{amssymb}	
\bibpunct{(}{)}{;}{a}{}{,}

\title[Globular clusters as tracers of the evolutionary history]{
Globular cluster systems as tracers of the evolutionary history in 
NGC\,3258 and NGC\,3268 \thanks{Based on observations
collected at the Cerro Tololo Interamerican Observatory (CTIO); observations 
carried out at the European Southern Observatory, Paranal (Chile), program 
71.B-0122(A), and observations obtained at the Gemini Observatory, which is 
operated by the Association of Universities for Research in Astronomy, Inc., 
under a 
cooperative agreement with the NSF on behalf of the Gemini partnership: the 
National Science Foundation (United States), the National Research Council 
(Canada), CONICYT (Chile), Ministerio de Ciencia, Tecnolog\'ia e Innovaci\'on 
Productiva (Argentina), and Ministerio da Ciencia, Tecnologia e Inovasao 
(Brazil).}}
\author[J. P. Caso et al.]
{Juan Pablo Caso$^{~1,2}$\thanks{E-mails:\,jpcaso@fcaglp.unlp.edu.ar\,(JPC);\,lbassino@fcaglp.unlp.edu.ar
(LPB);\,mastiasgomez@unab.cl\,(MGC)}, 
Lilia P. Bassino$^{~1,2}$ and Mat\'ias G\'omez$^{~3}$\\ 
$^{1}$Facultad de Ciencias Astron\'omicas y Geof\'isicas de la Universidad Nacional de     
La Plata, and \\ Instituto de Astrof\'isica de La Plata (CCT La Plata -- CONICET, UNLP), 
Paseo del Bosque S/N, B1900FWA La Plata, Argentina\\   
$^{2}$Consejo Nacional de Investigaciones Cient\'ificas y T\'ecnicas, Rivadavia 1917, 
C1033AAJ Ciudad Aut\'onoma de Buenos Aires, Argentina\\   
$^{3}$Departamento de Ciencias F\'isicas, Facultad de Ciencias Exactas, Universidad Andres Bello, 
Fernandez Concha 700, Las Condes, Chile} 

\date{Accepted XXX. Received YYY; in original form ZZZ}

\pubyear{2017}  

\begin{document}
\label{firstpage}
\pagerange{\pageref{firstpage}--\pageref{lastpage}}
\maketitle

\begin{abstract}
We present a new photometric study of NGC\,3258 and NGC\,3268 globular
cluster systems (GCSs), using images in filters $B,C,V,R,I$ and $z'$, 
obtained from four different telescopes. The wide spatial coverage  
allow us to estimate the whole extension of both GCSs more precisely 
than in previous works, and new values for the richness of GCs
subpopulations. We find differences in the azimuthal 
distribution between blue (metal-poor) and red (metal-rich) globular clusters (GCs), and confirm
that radial profiles flatten towards the centre of the galaxies. In both
cases we detected a radial gradient in the colour peak of blue GCs which 
might be related to the construction of the GCSs. We analyse the
similarities and differences in both GCSs, in the context of the posible 
evolutionary histories of the host galaxies. We also obtain 
photometric metallicities for a large number of GC candidates around 
NGC\,3258, by applying multicolour-metallicity relations. These results
confirm the bimodal metallicity distribution.
\end{abstract}


\section{Introduction}
\label{intro}
Despite some noteworthy cases, the bulk of the globular clusters systems 
(GCSs) are usually old stellar systems \citep[e.g.][]{bro06}, 
formed under environmental conditions achieved during massive star formation 
episodes \citep{ash92,lar00,kru14}. This fact implies a direct connection 
between the formation of GCSs and field stars populations, which might be useful 
to describe the evolutionary history of a galaxy based on the study of its GCS 
\citep[e.g.][]{cas13b,cas15a,esc15}.  

The most studied property of GCSs, mainly in early-type galaxies,
is the bimodality of their colour distributions. This behaviour is
usually understood as a result of bimodality in metallicity, in agreement with 
spectroscopic results \cite[e.g.][]{woo10b,ush12,can14}, despite
other interpretations have been proposed \citep[e.g.][]{yoo06,ric13}. 

In this sense, we can highlight two theories in the current scheme 
of GCSs origin. According to one of them, massive star formation episodes, consequence
of the merging of building blocks of proto-galaxies, are responsible for the 
formation of blue GCs ({\it bona fide} low-metal content) at high redshift, 
while the red ones ({\it bona fide} high-metal content) are formed in subsequent 
merger events, after gas enrichment by stellar evolution \citep{mur10,li14}.
Alternatively, \citet{ton13} suggests a hierarchical clustering model where the 
red GC subpopulation is composed by clusters formed in the galaxy main 
progenitor around redshift $z\approx 2$, while the blue GC subpopulation is
composed by clusters accreted from satellites, and formed at redshifts 
$z\approx 3-4$. Both scenarios can explain the age difference found between 
the two GC subpopulations in the Galaxy \citep{lea13,han13}.

Our target galaxies, NGC\,3258 and NGC\,3268, belong to the Antlia galaxy 
cluster, located in the Southern sky at a low Galactic latitude ($\approx 19\degr$).
Its galaxy content was originally studied by \citet{hop85} and 
\citet{fer90}, and more recently by \citet{smi12} and \citet{cal15}. 
The central part consists of two subgroups, each one dominated  
by one of these giant elliptical (gE) galaxies (i.e. NGC\,3258 and NGC\,3268) 
of almost the same luminosity. The cluster might be in a merging process, 
but SBF distances \citep{bla01,can05,tul13} and radial velocities analysis 
\citep{hes15,cas15b} are not conclusive.

The GCSs around the two gEs were first analysed by 
\citet{dir03b}. Afterwards, studies of the inner region of both GCSs were 
carried out, with deeper photometry, by \citet{bas08} with VTL data and 
\citet{har06,har09a} with Hubble Space Telescope (HST) data.

\medskip
Our goal is to enhance previous studies by taking advantge of wider and 
deeper datasets. From these we have derived robust photometric metallicities 
for a large sample of GCs in NGC\,3258 for the first time.

This paper is organized as follows. The observations and data reduction are 
described in section 2, and the results are presented in section 3, while 
section 4 is devoted to the discussion. Section 5 summarizes the concluding 
remarks.

\section[]{Observations and data reduction}

\subsection{MOSAIC data}
A subset of the data consists of wide-field images in the Washington photometric 
system, taken with the MOSAIC\,II camera mounted on the  CTIO 4-m Blanco telescope. 
One field corresponds to the central region of the Antlia cluster while another 
field is located to the East (hereafter, CF and EF). The observations were performed 
during 4/5 April 2002 in the case of the CF, and during 24/25 March 2004 for the EF. 

\begin{figure}
 \includegraphics[width=80mm]{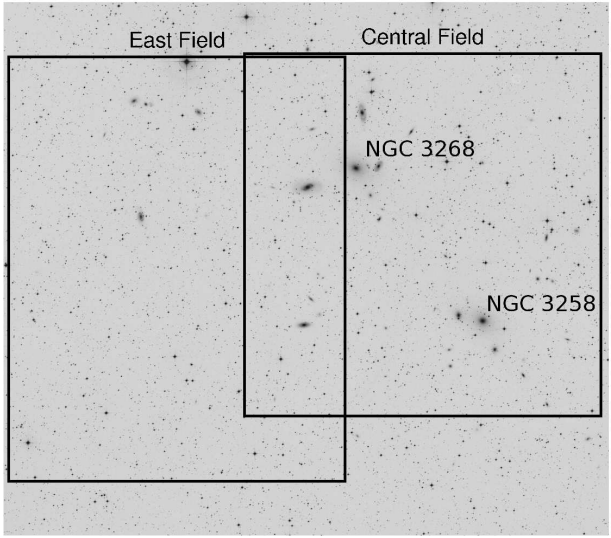}
 \caption{The MOSAIC fields are overlaid on a $65 \times 65$\,arcmin$^2$ DSS image 
of the Antlia cluster. North is up, East to the left.}
 \label{dss}
\end{figure}

For the CF, four images in $R$ and seven in $C$ were obtained, all of them with 
an exposure time of 600\,s. For the EF five 600\,s images in $R$ and seven
900\,s exposure in $C$ were observed. The Figure\,\ref{dss} shows the location 
of both MOSAIC fields, overlaid on a DSS image.

We used the Kron-Cousins $R$ and Washington $C$ filters, although the genuine 
Washington system uses $T_1$ instead of $R$. However, \citet{gei96a} has shown 
that the Kron-Cousins $R$ and $T_1$ magnitudes only differ in the zero-point 
(we used $R-T_1=-0.02$ from \citealt{dir03b}). The data were dithered in order 
to fill in the gaps between the eight individual CCD chips. The MOSAIC pixel
scale is $0.24$\,arcsec/pixel and its field-of-view (FOV) is $36\times36$\,arcmin$^2$.
Different FOV between CF and EF in Figure\,\ref{dss} are due to a large 
dithering in the 2004 observations, nedeed to solve an amplifier failure.

\subsubsection{Data reduction}

The MOSAIC data has been handled using the MSCRED package within {\sc IRAF}. 
In order to facilitate point sources detection, the extended galaxy light was 
subtracted, using a ring median filter with an inner radius of 9\,arcmin and 
an outer radius of 11\,arcmin. 

We applied the software SE\textsc{xtractor} \citep{ber96} to the $R$ images, in 
order to obtain an initial selection of point sources. The software considered 
as a positive detection every group of five or more connected pixels with 
counts above $1.5\times\sigma$ the sky level. The effective radius of classic 
GCs usually does not exceed a few parsecs \citep[e.g.,][]{har09a}. This implies 
that, at the adopted Antlia distance of $\approx35$\,Mpc 
\citep[distance modulus $m-M = 32.73$,][]{dir03b}, GCs are seen as point 
sources on our MOSAIC images. Thus, we used the star/galaxy classifier from 
SExtractor, to generate the point sources' catalogue.
We performed the photometry with the DAOPHOT package \citep{ste87}. For
both filters, a spatially variable PSF was modelled, employing about a hundred 
bright stars, well distributed over each field. The final point source 
selection was based on the $\chi$ and sharpness parameters of the ALLSTAR 
task.

In the case of the CF, we used the calibration equations from \citet{dir03b}. 
For the EF, the calibration equations were obtained from standard star fields 
observed during both nights of the observing run. For each night, 4 to 5 fields, 
containing about 10 standard stars from the list of \citet{gei96b} were observed, 
spanning a large range in airmass (typically from 1.0 to 1.9). The fitted 
coefficients for each nights were indistinguishable within the uncertainties, and 
hence we used a single set of transformation equations. 

Finally, the equations for the EF are:
\\
\begin{eqnarray}
  (C-T_1)& = & (c-r) - 0.687 - (0.418 \times X_C \\
        & + & -0.14 \times X_R) + 0.092 \times (C-T_1) \nonumber \\
     T_1 & = & r + 0.628 - 0.14 \times X_R \\ 
        & + & 0.019 \times (C-T_1) \nonumber
\end{eqnarray}

\noindent where $X_C$ and $X_R$ are the mean airmass coefficients for each filter. 
While $(c-r)$ and $r$ are the instrumental colour and magnitude, $(C-T_1)$ and $T_1$ 
are the calibrated ones.

Aperture corrections were obtained from the stars selected for each PSF, 
and extinction corrections were calculated from \citet{sch11} values (available
in NED\footnote{This research has made use of the NASA/IPAC 
Extragalactic Database (NED) which is operated by the Jet Propulsion 
Laboratory, California Institute of Technology, under contract with the 
National Aeronautics and Space Administration.})
The colour excess for $(C-T_1)$ was calculated as $E_{(C-T_1)} = 1.97 \times E_{(B-V)}$ 
\citep{har77}.

As both fields partially overlap (Fig.\,\ref{dss}), we have been able to determine 
zero point differences between them from common point sources.
Differences in $C$ filter are marginal, $-0.012\pm0.004$, but $T_1$ 
magnitudes in the EF are $0.121\pm0.002$\,mag fainter than CF ones.
We compared the $(C-T_1)$ colours of red Galactic stars and blue background 
galaxies from the CF photometry, easily identified in the colour-magnitude diagram 
(CMD), with those available in the literature in the same photometric system, e.g.   
for the GCSs of NGC\,1399 \citep{bas06a}, NGC\,4636 \citep{dir05}, and M87 \citep{for07}. 
In all cases, the colours of Galactic stars and background galaxies agree with 
those corresponding to our EF field. Hence, we decided to refer the photometry to this 
latter field and apply the zero-point corrections to the CF one.

\subsubsection{Photometric completeness}

The data completeness has been studied with the aid of the task ADDSTAR 
within IRAF, adding 1\,000 artificial stars to each science image. These
stars present an homogeneous spatial coverage, and span the typical ranges 
in $T_1$ magnitude and $(C-T_1)$ colour as GCs. Then, photometry was
carried in the same manner as the original data. This was repeated ten 
times, resulting in a total of 10\,000 artificial stars, and the results
were grouped in a single catalogue.

\begin{figure}
 \includegraphics[width=43mm,angle=270]{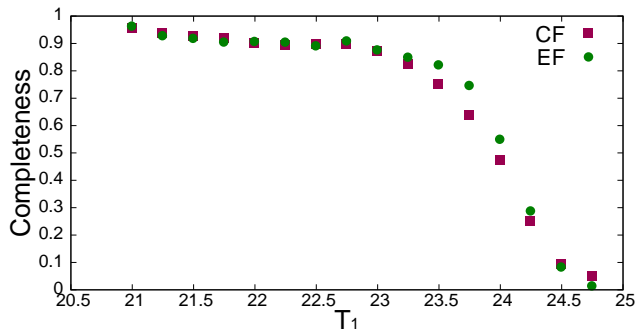}
 \caption{Completeness curves for the MOSAIC fields.} 
 \label{complmos}
\end{figure}

Each field was divided in nine regions, in order to search for strong 
spatial variations, but no trends were detected, with the exception
of the gEs inner region. For these reason, all the point sources at 
less than 1\,arcmin from gEs centre were avoided. The completeness 
functions in both MOSAIC fields are similar (Figure\,\ref{complmos}). 
The completeness in the fields falls below $90\,\%$ at $T_{1,0} \approx 
22.6$\,mag, and the $60\,\%$ completeness is reached at $T_{1,0} \approx 
23.85$\,mag.

Finally a single photometric catalogue was built with the point sources from 
both fields. In the overlapping region priority was given to the CF photometry.

\subsubsection{Selection criteria}
\label{mos.sc}

GC candidates were selected as points sources with $0.9 < (C-T_1)_0 < 2.3$\,mag and 
$M_V > -10.5$. The colour range proposed here is similar to those used in literature 
\citep[e.g.][]{dir03a,dir05,bas06a}.
The magnitude limit, which implies $V \approx 22.23$\,mag at Antlia distance, was 
adopted to avoid ultra-compact dwarf (UCD) candidates \citep[e.g.][]{mie12,nor14},
 previously studied in \citet{cas13a} and 
\citet{cas14}. Assuming the mean $(V-R)$ colour for Es from 
\citet{fuk95} to transform this latter magnitude, to select GCs candidates we 
chose $T_{1,0}=21.6$\,mag as the bright end and $T_{1,0}= 23.75$\,mag as the faint one.

The background region selected to correct for the contamination, in the MOSAIC 
photometry, covers an area of 489\,arcmin$^2$ and is located at more than 25\,arcmin 
from the gEs, in the Eastern portion of the EF. At this distance, the presence of 
GCs from NGC\,3268 system should be negligible, considering that the extrapolation
of the GC radial distribution at this galactocentric distance (see Section\,\ref{radsec})
is less than a seventh of the mean density in the region.

\subsection{FORS1 data}

We also used FORS1--VLT images in the $V$ and $I$ bands (program 71.B-0122(A),   
PI B. Dirsch). These images correspond to three fields, two of them centred 
on each one of the gEs, and the third one located to the North-West direction 
(see Figure\,1, from \citealt{bas08}, hereafter N3258F, N3268F and BF,
respectively). Their pixel scale is 0.2\,arcsec/pixel and their FOV of
$6.8\times6.8$\,arcmin$^2$.
We refer to \citet{bas08} and \citet{cas13a} for further details 
on the reduction and photometry of these data.

\subsubsection{Photometric completeness}

\begin{figure}
 \includegraphics[width=43mm,angle=270]{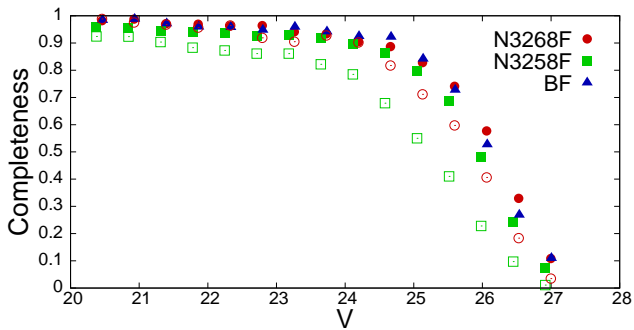}
 \caption{Completeness curves for the three FORS1 fields. For N3258F and N3268F 
we discriminated between objects at less than 1\,arcmin from galaxy centre (open 
symbols) and out of this limit (filled symbols).} 
 \label{complvlt}
\end{figure}

The procedure was similar to that applied to the MOSAIC photometry. 
This time we added 500 artificial stars to each image, and repeated 
this procedure 40 times, totalizing 20\,000 stars. The results are shown 
in Figure\,\ref{complvlt}, where the completeness curves for N3258F 
and N3268F were determined independently for objects at less than 1\,arcmin 
from galaxy centre (open symbols) and out of this limit (filled symbols). 
Completeness curves for galactocentric distances larger than 1\,arcmin are 
similar in the three fields, achieving $50\%$ at $V_0=26$\,mag. 

\subsubsection{Selection criteria}

Following similar criteria as those described in Section\,\ref{mos.sc}, in this 
case the GC candidates were selected with $0.6 < {\bf (V-I)_0} < 1.4$\,mag and 
$25.5 < V_0 < 22.2$\,mag. The BF was used to estimate the contamination. 

\subsection{ACS data}

In addition, images of an ACS field on NGC\,3258 were obtained from the Hubble 
Space Telescope Data Archive\footnote{Some of the data presented in this 
paper were obtained from the Mikulski Archive for Space Telescopes (MAST). 
STScI is operated by the Association of Universities for Research in 
Astronomy, Inc., under NASA contract NAS5-26555. Support for MAST for 
non-HST data is provided by the NASA Office of Space Science via grant 
NNX09AF08G and by other grants and contracts.} (programme 9427, PI: W. E. 
Harris), which will be used just to estimate photometric metallicities. 
The resulting images are the composite of four 1340\,s exposures 
in the $F435$ filter and four 570\,s in the $F814$ one, where individual 
exposures were observed in October 2002. Moreover, a field of 47\,Tucanae 
outskirts was used to model the PSF. These observations had been carried out
in the same filters as NGC\,3258, with exposure times of $30s$, in September 
2002 (programme 9656).
These images present a pixel scale of 0.05\,arcsec/pixel and a FOV of 
$202\times202$\,arcsec$^2$.

\subsubsection{Photometry}

With HST GC-like objects might be marginally resolved at NGC\,3258 distance 
\citep[e.g.][]{cas13a,cas14}. Hence, we ran SE\textsc{xtractor} on 
the images in both filters, and considered as likely GC candidates 
those sources with ellipticity $\epsilon < 2$ and $FWHM < 5$\,px, 
similar to other GC studies developed with ACS images 
\citep[e.g.][]{jor04,jor07}. We performed aperture photometry with 
the PHOT task within \textsc{IRAF}, using an aperture radius of 5\,px.
In order to obtain the PSF from the 47\,Tucanae fields, approximately 
40 bright stars were used in each filter. Then, the \textsc{ISHAPE} 
\citep{lar99} software was used, considering such PSF, to calculate 
structural parameters for bright GC candidates, resulting in a typical 
effective radius $R_{\rm eff}$ of 0.33\,px (i.e., $R_{\rm eff} \approx 3$\,pc 
at Antlia distance). Approximately 20 GC candidates brighter than 
$F814W=23.5$\,mag, relatively isolated and with $0.28<R_{\rm eff} {\rm [px]} <0.38$ 
in both filters were used to calculate aperture corrections, resulting 
in $-0.09$\,mag for $F435W$ and $-0.10$\,mag for $F814W$.  

Calibrated magnitudes in $B$ and $I$ filters were obtained using the 
relation

\begin{equation}
m_{std} = m_{inst} + ZP \nonumber
\end{equation}

\noindent with zero points $ZP_{435}= 25.779$ and $ZP_{814}= 25.501$,
taken from \citet{sir05}.

Finally, we applied the \citet{sch11} Galactic extinction corrections  
obtained from NED.

\subsection{GMOS data}

We also used $z'$ images of a GMOS-Gemini South field, obtained during 
semester 2016A (programme GS-2016A-Q-66, PI: J.P. Caso). This field 
contains the gE NGC\,3258 (see Fig\,\ref{dss}) and was
observed as $22\times 200\,s$ exposures, slightly dithered in
order to fill in the gaps of the GMOS-S field and to efficiently 
remove cosmic rays and bad pixels. The pixel scale in GMOS field is
0.16\,arcsec/pixel and the FOV $5.5\times5.5$\,arcmin$^2$.

\subsubsection{Data reduction and photometry}

The image reduction was performed following standard procedures with 
the GEMINI package within \textsc{IRAF}.

First, we subtracted the extended galaxy light applying a median
filter. Then, we used SE\textsc{xtractor} in order to obtain an 
initial point sources' catalogue. The photometry was performed with 
the DAOPHOT package \citep{ste87} within \textsc{iraf}. A 
second-order variable PSF was generated from a sample of bright 
stars, well distributed over the field. This PSF was used to
calculate the PSF photometry with the ALLSTAR task. The final point 
source selection was made with the $\chi^2$ and sharpness 
parameters of ALLSTAR.

Aperture correction was obtained from the same bright and moderately 
isolated stars used to model the PSF and resulted $-0.23$\,mag.

\subsubsection{Photometric calibration}

A photometric standard stars field from the list of \citet{smi02} was 
observed together with the science data and reduced in the same manner.
We obtained aperture photometry for the stars located in the field in order
to fit transformation equations of the form:

\begin{equation}
z'_{std} = ZP + z'_{inst} - K_{CP} \times (X-1) + CT \times (i' - z') \nonumber
\end{equation}

\noindent where $m_{std}$ and $m_{inst}$ are the calibrated and instrumental 
magnitudes, respectively, and $ZP$ is the photometric zero point, which 
resulted $28.08\pm0.02$. $K_{CP}$ is the mean atmospheric extinction 
at Cerro Pach\'on, obtained from the Gemini Observatory Web 
Page\footnote{http://www.gemini.edu/sciops/instruments/gmos/calibration}, 
 $X$ is the airmass, and $CT$ the colour term, which is a minor correction
for GMOS observations ($CT= -0.04$ for $z'$ filter). As we do not 
have $i'$ observations for the standard stars, we obtained their $(i'-z')$ 
colour from the \citet{smi02} catalogue.

Afterwards, we applied the \citet{sch11} Galactic extinction corrections  
obtained from NED.

GMOS data will be only  used for photometric metallicities estimations in 
Section\,3.5, and completeness curves will not be applied. Despite of this,
we analyzed the quality of the photometry adding, 5\,000 artificial stars split 
in 20 images. Then we repeated the photometry in the same manner than the original
and compared the measured magnitudes for this set of artificial stars with the
input ones. The results showed that magnitudes are not affected neither by the
reduction nor photometry procedures.

\section{Results}

\begin{figure*}
\includegraphics[width=170mm]{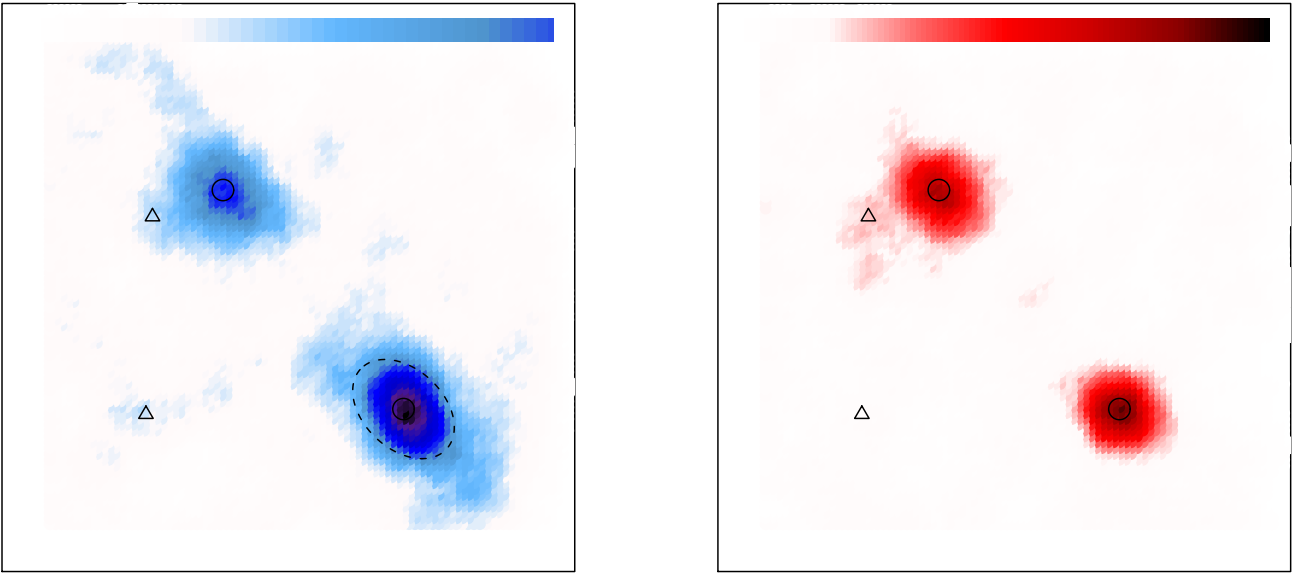}
\caption{The left panel shows the spatial distribution for blue GC candidates around 
NGC\,3268 (upper left) and NGC\,3258 (lower right) from the MOSAIC data. 
Both gEs centres are indicated 
with circles, while small triangles refer to the centre of NGC\,3273 and NGC\,3271, 
two disc galaxies with small GCSs. The dashed ellipse represents the parameters of
the spatial distribution derived from the azimuthal distribution (see text in 
Section\,\ref{espasec} and Figure\,\ref{acim}). The right panel is analogue but for 
red GC candidates. North is up and East to the left.}
\label{espa}
\end{figure*}

\subsection{Spatial distributions}
\label{espasec}

The left panel of Figure\,\ref{espa} shows the spatial distribution of GC candidates 
around NGC\,3268 and NGC\,3258 from MOSAIC data for the colour range 
$0.9<(C-T_1)_0<1.5$, assumed as the blue
subpopulation. The figure spans $40.5\times 40.5$\,arcmin$^2$, and
represents the smoothed counts, obtained after dividing the region in $0.5$\,arcmin-side cells. 
The right panel is analogue to the previous one, but devoted to the red GCs, which are 
assumed to span $1.5<(C-T_1)_0<2.2$. Hereafter, the terms blue and red GCs will refer to 
these colour ranges.

The projected distribution of blue GCs around NGC\,3258 is elongated, in a direction 
that roughly matches the one that crosses both galaxy centres. In the case of NGC\,3268,
the elongation is not so evident, but an overdensity of GCs towards NGC\,3258 seems to 
be present. For both galaxies, the spatial distribution of red GCs is nearly circular, 
though that of NGC\,3268 looks slightly elongated in the same direction. \citet{dir03b} 
and \citet{bas08} indicated that the projected distribution of ``all'' GCs in NGC\,3258 
and NGC\,3268 seem to be elongated in the direction towards the other gE. From the 
results in Fig.\,\ref{espa}, it is worth discriminating between blue and red GCs in both 
galaxies, assuming the colour ranges previously indicated. As both GCSs are 
dominated by blue GCs, the azimuthal distributions for red ones are represented 
with a larger step ($40\degr$ instead of $30\degr$).

From FORS1 data, a similar behaviour in the spatial distribution was
found for both galaxies in the limited radial range where azimuthal completeness
is achieved.

\begin{figure}
\includegraphics[width=84mm]{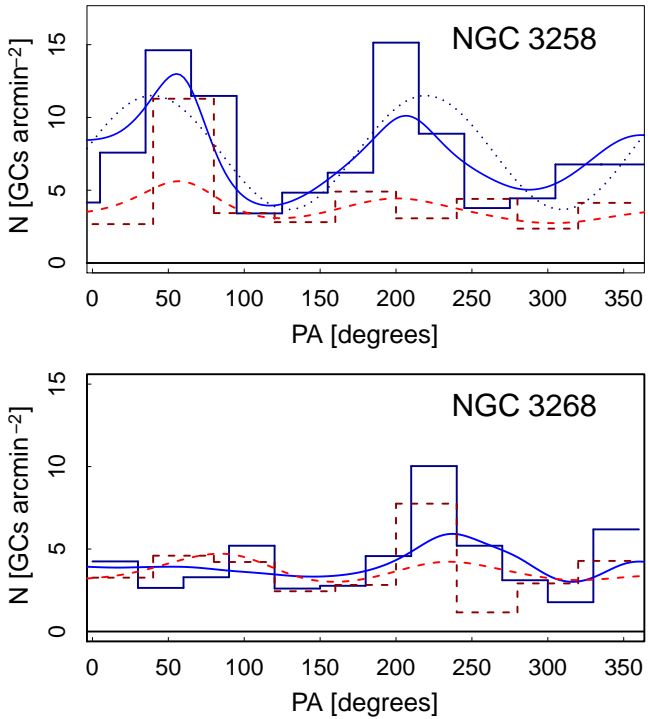}
\caption{Azimuthal distribution of GC candidates within 4\,arcmin from the 
centre of NGC\,3258 (upper panel) and NGC\,3268 (lower panel). In both cases, 
we discriminate between blue (solid histogram) and red GCs (dashed
histogram). The solid and dashed curves indicate the smoothed distributions
for each subpopulation, obtained with a Gaussian kernel. The dotted curve in 
the upper panel shows the least-squares fitting to the blue GCs distribution.}
\label{acim}
\end{figure}

The upper panel of Figure\,\ref{acim} shows the azimuthal distribution for NGC\,3258 
GC candidates located within 4\,arcmin from the galaxy centre, as the number of GCs versus 
position angle ($PA$). We select this radial limit due the presence of several bright 
galaxies close to both gEs in projected distance. The $PA$ was measured from North, to
the East. We separated blue and red subpopulations, represented by solid and dashed
histograms, respectively. The solid and dashed curves indicate the smoothed distributions 
for each subpopulation, obtained with a Gaussian kernel. 
The azimuthal distribution for red GCs shows
a single maximum at $PA \approx 60\degr$, similar to the projected direction to 
NGC\,3268. Blue GC distribution presents a peak at $\approx 50\degr$, together with 
a second one separated by $\approx 150\degr$. These results are similar to what is 
expected for elongated distributions, taking into account the adopted step. According 
to this, the dotted line indicates the sinusoidal curve with a $\pi$-period fitted 
to the distribution. Its amplitude and PA are $4.2\pm1.2$ and $42.5\pm3\degr$.

The lower panel refers to NGC\,3268 GCS. 
For both subpopulations, the spatial distribution presents a clearly larger number of 
GCs around $PA \approx 220\degr$, which matches with the projected direction to 
NGC\,3258. The smoothed distribution for red GCs seems to present a second maximum at
 $\approx 80\degr$, which might indicate an elongated distribution, but the evidence 
is not conclusive.

There is plenty of evidence in the literature that points to a close relation 
between red GCs and galaxy starlight in ellipticals, while blue GCs probably follow 
the total mass distribution \citep[e.g.][and references therein]{for12}. In 
this context,
our results are in agreement with the parameters of the diffuse starlight in both 
gEs, that present some ellipticity for the inner 2.5\,arcmin and a PA of $\approx 70^o$
\citep[e.g.][]{dir03b,bas08}.
The diffuse X-ray emission in both galaxies are also elongated \citep{ped97,nak00},
something that could link blue GC distributions to them.

\subsection{Radial distributions}
\label{radsec}

While MOSAIC spatial coverage allow us to determine the total extension of the GCS, 
as well as an improved background correction, the depth of FORS1 data results
in accurate statistics for the inner part of the GCSs. Therefore, we complement the
results from both photometries in order to obtain the radial profiles for blue and 
red GCs.

From the MOSAIC data we determined the radial distribution for galactocentric 
distances larger than 1\,arcmin. In order to avoid changes in the completeness due 
to the presence of the three NGC\,3268 bright neighbours, we excluded GC candidates 
near the other bright galaxies in the field (i.e., at less than 1\,arcmin from 
NGC\,3257, NGC\,3260 NGC\,3267 or NGC\,3273, and 3\,arcmin from NGC\,3271). In 
addition, the gEs GCSs probably overlap as suggested by \citet{bas08}. As a 
consequence, we obtained the extension of both GCS in an iterative way, avoiding 
candidates in regions where previous studies indicated that a contribution from the 
other GCS was expected. For both gEs we proposed that the total extension for the 
GCS is achieved when the GC surface density is equal to 30\,per\,cent of the background 
level. This criterium has been successfully applied to the study of other GCSs 
\citep[e.g.][]{cas15a,bas17}, including those developed with MOSAIC images 
\citep[e.g.][]{bas06a,cas13b}. The radial profiles were corrected for contamination 
and the errors were assumed as Poissonian.

\begin{figure}
\includegraphics[width=84mm]{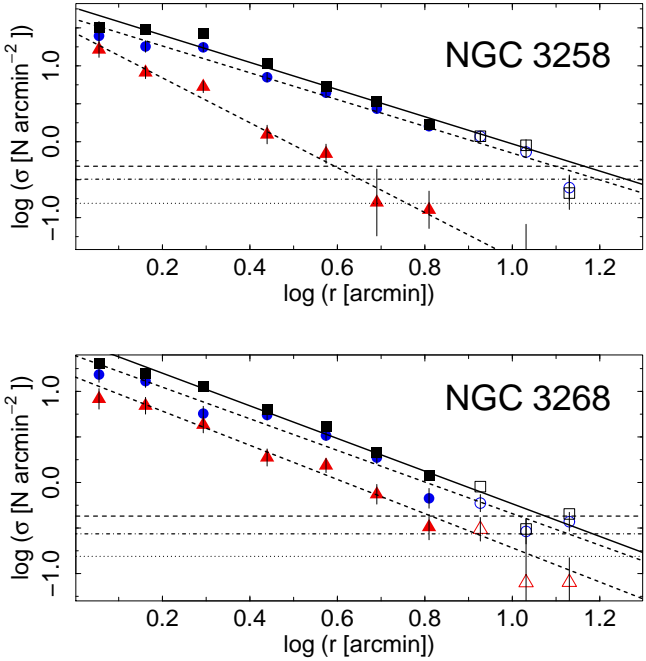}
\caption{Logarithmic radial density profiles for GC candidates around 
NGC\,3258 (upper panel) and NGC\,3268 (lower panel) from MOSAIC
data. Squares represent
the entire sample, circles indicate blue GCs, while triangles show red 
ones. The horizontal lines refer to the $30\,\%$ of the background 
level for the entire sample (dashed), the blue GCs (dot-dashed) and
the red ones (dotted). This limit is proposed to estimate the GCS extension. 
The three profiles are background corrected.}
\label{rad}
\end{figure}

The upper panel of Figure\,\ref{rad} corresponds to the radial projected 
distribution of GCs around NGC\,3258 derived from MOSAIC data.
Galactocentric distances smaller than 1\,arcmin are avoided 
due increasing incompleteness. Radial bins that do not overlap with NGC\,3268 GCS (from the
iterative process, those with projected distance to NGC\,3258 lower than $r=7$\,arcmin)
 are indicated with filled symbols. Open symbols represent the projected density at 
galactocentric radii where both GCS might be overimposed, i.e. the ones where we avoid 
the overlapping region, which substantially reduces the analyzed area. For blue GCs, we 
obtain the radial distribution using concentric ellipses with position angle PA=$41\degr$ 
(Figure\,\ref{acim}) and ellipticities $\epsilon=0.32$, calculated from our results
as $\epsilon= 1-(N_b/N_a)^{1/\alpha}$ \citep[][$N_a$ and $N_b$ being the number of GCs along
major and minor axes, and $\alpha$ the slope absolute value for the radial profile when
azimuthal symmetry is assumed]{dir03b}. As an example, an ellipse with this parameters
and major axis 4.5\,arcmin is plotted in Figure.\,\ref{espa}. 
The dashed horizontal line represents 30\,per\,cent of the background level for the entire 
sample, while dot-dashed and dotted lines correspond to blue and red GCs, respectively. 
We indicate the results of fitting a power-law to blue (circles) and red (triangles) GC 
distributions with dashed lines, while the solid one represents the power-law
fitted to the entire sample of GC candidates (squares). The extrapolation in both 
curves indicates an extension for the GCS of $\approx 17$\,arcmin
($\approx$\,170\,kpc at Antlia distance) that is defined by the blue subpopuation, 
while the red subpopulation seems to be
more concentrated towards NGC\,3258, reaching 6\,arcmin . The slopes for the fitted
power-laws were $-1.76\pm0.10$ and $-2.90\pm0.15$ for the blue and red GCs, respectively.

The lower panel corresponds to the radial projected distribution of GCs around 
NGC\,3268 from MOSAIC data, for galactocentric distances larger than 1\,arcmin.
In this case, we calculated the 
radial distributions using concentric circles, considering that the evidence of 
an elliptical distribution is not so conclusive as in NGC\,3258. The extension 
for the blue and red subpopulations are very similar, $\approx 14$ and 
$\approx 12$\,arcmin, respectively. In this case the fitted slopes for blue and 
red GC candidates were $-1.70\pm0.13$ and $-1.80\pm0.15$, respectively.

Several studies point to a flatter radial profile in the inner regions of 
GCSs \citep[e.g.][]{els98}, probably due to GC erosion during early stages
of the system evolution \citep[e.g.][]{cap09,bro14}. In order to extend 
the radial profiles from MOSAIC data to these innermost regions, we also 
obtained the radial profiles from the VLT data. Unlike previous analysis 
from \citet{bas08},
we corrected the profiles by completeness considering different curves 
for bins within 1\,arcmin or outside this limit (see Figure\,\ref{complvlt}). 
The background correction was obtained from a field
located to the North-West of NGC\,3268 \citep[see Figure\,1 from][]{bas08}.
This field might present a small number of GCs from NGC\,3268 system, but
surface density at large galactocentric distances is very low, and hence
contamination from NGC\,3268 clusters should be negligible. 

In order to fit the FORS1 profile, we adopted a modified Hubble distribution 
\citep{bin87,dir03b},

\begin{equation}
n(r) = a\left( 1+ \left(\frac{r}{r_0}\right)^2\right)^{-\beta}
\end{equation}

\noindent which behaves as a power-law with slope $2\beta$ for large radii.
Despite its limited photometric depth, the large FOV of MOSAIC images 
allow us to obtain a more accurate determination of the slope of radial 
distributions than FORS1 fields.
Thus, for fitting the Hubble distribution we fixed $\beta$ as a half the 
slope of the corresponding power-law obtained from MOSAIC data. 
Hence, only $a$ and $r0$ remained as free parameters to fit.

\begin{figure}
\includegraphics[width=84mm]{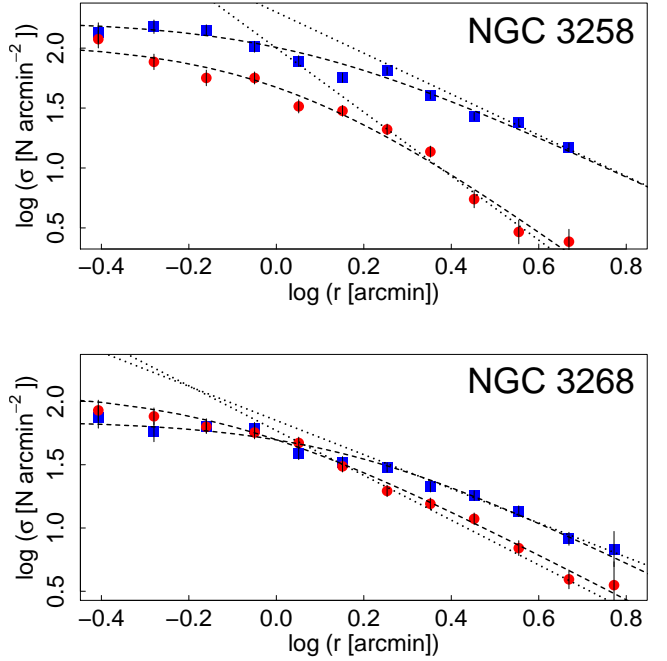}
\caption{Background and completeness corrected radial density profiles 
from FORS1 data
for GC candidates around NGC\,3258 (upper panel) and NGC\,3268 (lower
panel) brighter than $V_0=25.5$, expressed in logarithmic scale. Circles 
represent blue GCs, while triangles indicate red ones. Dotted lines are 
power-law distributions obtained for the outer GCS from MOSAIC data, 
while dashed curves are the modified Hubble distributions fitted to these
samples.}
\label{radVI}
\end{figure}

The upper panel of Figure\,\ref{radVI} shows the radial profile for GCs 
around NGC\,3258 brighter than $V_0=25.5$, background and completeness 
corrected. We discriminated 
in blue (squares) and red (circles) subpopulations using the colour limit 
$(V-I)_0=1.05$, which has been previously used by \citet{bas08}. The 
dotted curves represent the power-laws previously fitted for the outer GCS 
region from MOSAIC data. It can be noticed the good agreement between
the fitted power-laws and the FORS1 data.
Dashed lines represent the modified Hubble distributions fitted to 
FORS1 data. The lower panel is analogue, but for NGC\,3268. Table\,\ref{hub}
shows the parameters obtained for both distributions.

\begin{table}   
\begin{center}   
\caption{Parameters of the modified Hubble distribution fitted to the FORS1 
radial profiles.}    
\label{hub}   
\begin{tabular}{@{}ccccc@{}}   
\hline   
\multicolumn{1}{@{}c@{}}{}&\multicolumn{2}{c@{}}{NGC\,3258}&\multicolumn{2}{c@{}}{NGC\,3268}\\   
\hline
\multicolumn{1}{c@{}}{}&\multicolumn{1}{c@{}}{Blue GCs}&\multicolumn{1}{c@{}}{Red GCs}&\multicolumn{1}{c@{}}{Blue GCs}&\multicolumn{1}{c@{}}{Red GCs}\\   
\hline
$a$&$2.23\pm0.05$&$2.05\pm0.06$&$1.85\pm0.03$&$2.09\pm0.06$\\
$r_0$&$1.11\pm0.10$&$1.05\pm0.08$&$1.40\pm0.10$&$0.77\pm0.08$\\
\hline
\end{tabular}    
\end{center}    
\end{table}

It can be clearly seen that the modified Hubble distribution provides an 
excellent fit to the inner regions for both GC subpopulations. 

\subsection{Luminosity functions and size of the GCSs}

FORS1 data were previously used by \citet{bas08} to obtain the GC luminosity 
function (GCLF) and distance through the turn-over magnitudes (TOM). We
repeated the measurement, taking into account the different completness curve 
for point sources within 1\,arcmin from the galaxy centre and considering a 
magnitude limit $V_0=25.5$. As this latter magnitude limit is barely fainter 
than the TOM obtained by \citet{bas08}, it might result in large uncertainties
in our TOM determination. In order to avoid this, we fixed the TOM to the 
distance moduli published by \citet{tul13}, $m-M= 32.56\pm0.14$ for NGC\,3258
and $m-M= 32.74\pm0.14$ for NGC\,3268. We have chosen these distances
because they are the most recent and accurate SBF estimations.

\begin{figure}
\includegraphics[width=84mm]{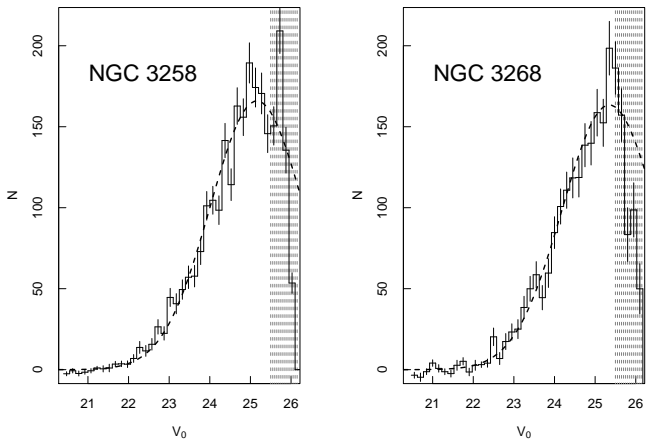}
\caption{GCLF for both gEs in the radial range $0.5 < R_g < 4.5$\,arcmin. Vertical 
grey lines indicate magnitudes fainter than the limit $V_0=25.5$. The dashed 
curves indicate the fitted Gaussian profiles, assuming the TOM derived from 
the respective \citet{tul13} distance modulus.}
\label{GCLF}
\end{figure}

In Figure\,\ref{GCLF} we present the GCLF for both gEs in the radial range 
$0.5 < R_g < 4.5$\,arcmin, where vertical grey lines indicate regions fainter than the
magnitude limit, and dashed curves show the fitted functions. 
We assume that both GCLF are well represented by Gaussian profiles with 
a nearly universal TOM $M_V= -7.4$\,mag \citep[e.g.][]{bro06,jor07}, resulting
in dispersions $\sigma_V= 1.16\pm0.04$ for NGC\,3258 and $\sigma_V= 1.15\pm0.04$
for NGC\,3268. From numerical integrations, we calculate that the GCs brighter than 
$V_0=25.5$ represent $62\pm6$ and $56\pm5$\,per\,cent, respectively, of the total samples. 
The errors are dominated by the distance moduli uncertainties. 

Hence, we can now estimate new GC populations for NGC\,3258 and NGC\,3268, considering 
a more precise completeness correction and a better fit of the radial GC distributions. 
From the numerical integration of the modified Hubble distributions fitted to the radial 
profiles (Section\,\ref{radsec}), we can calculate the GC population brighter than
$V=25.5$ in both cases. In order to obtain uncertainties, we calculated 1\,000 artificial
sets of Hubble distribution parameters using Monte-Carlo simulations, assuming that they
are generated by normal distributions with dispersions equal to the fitting errors. Then,
the deviation estimated from their numerical integration is assumed as the uncertainty 
in the GC population.

\begin{figure*}
\includegraphics[width=80mm]{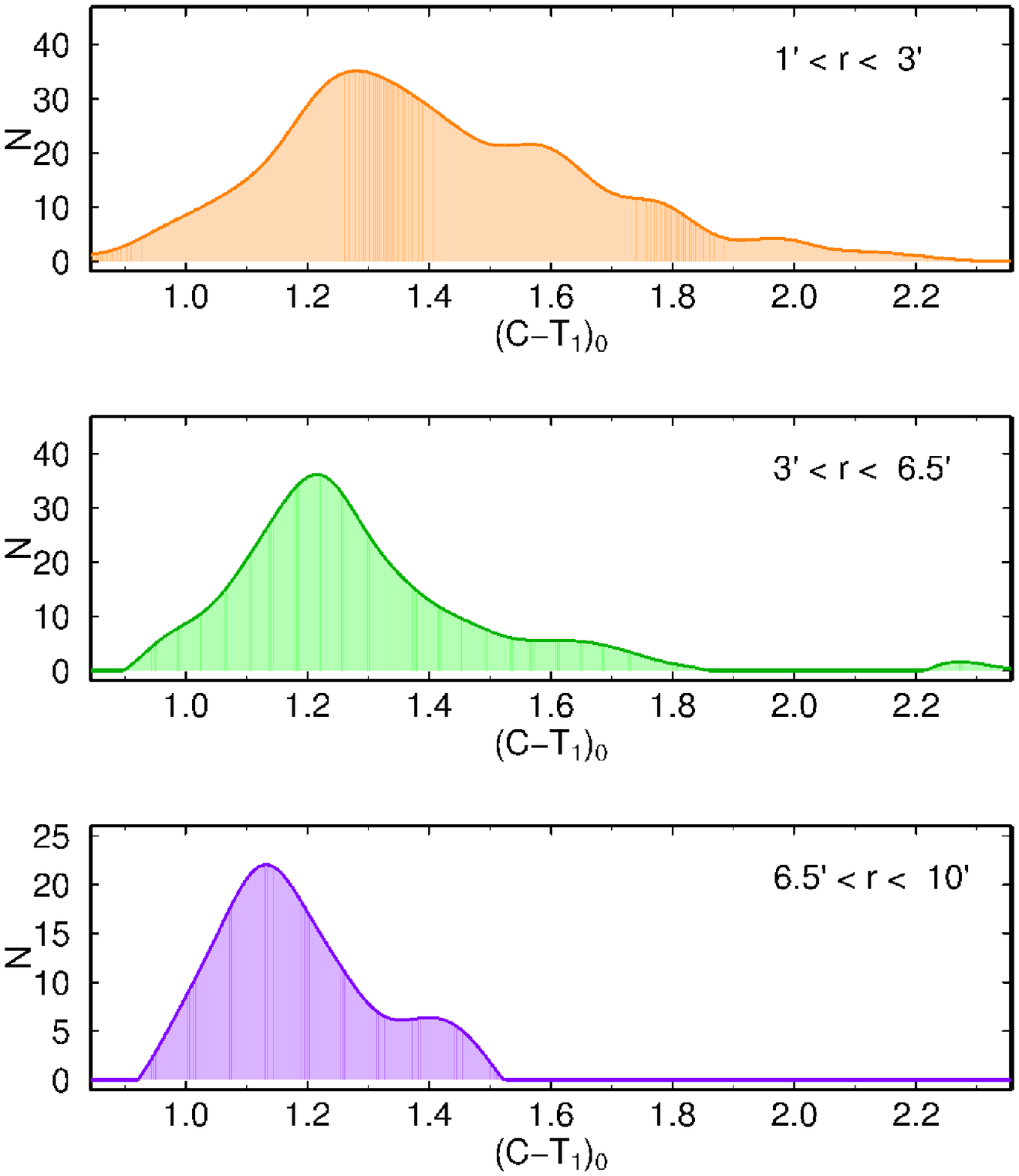}
\includegraphics[width=80mm]{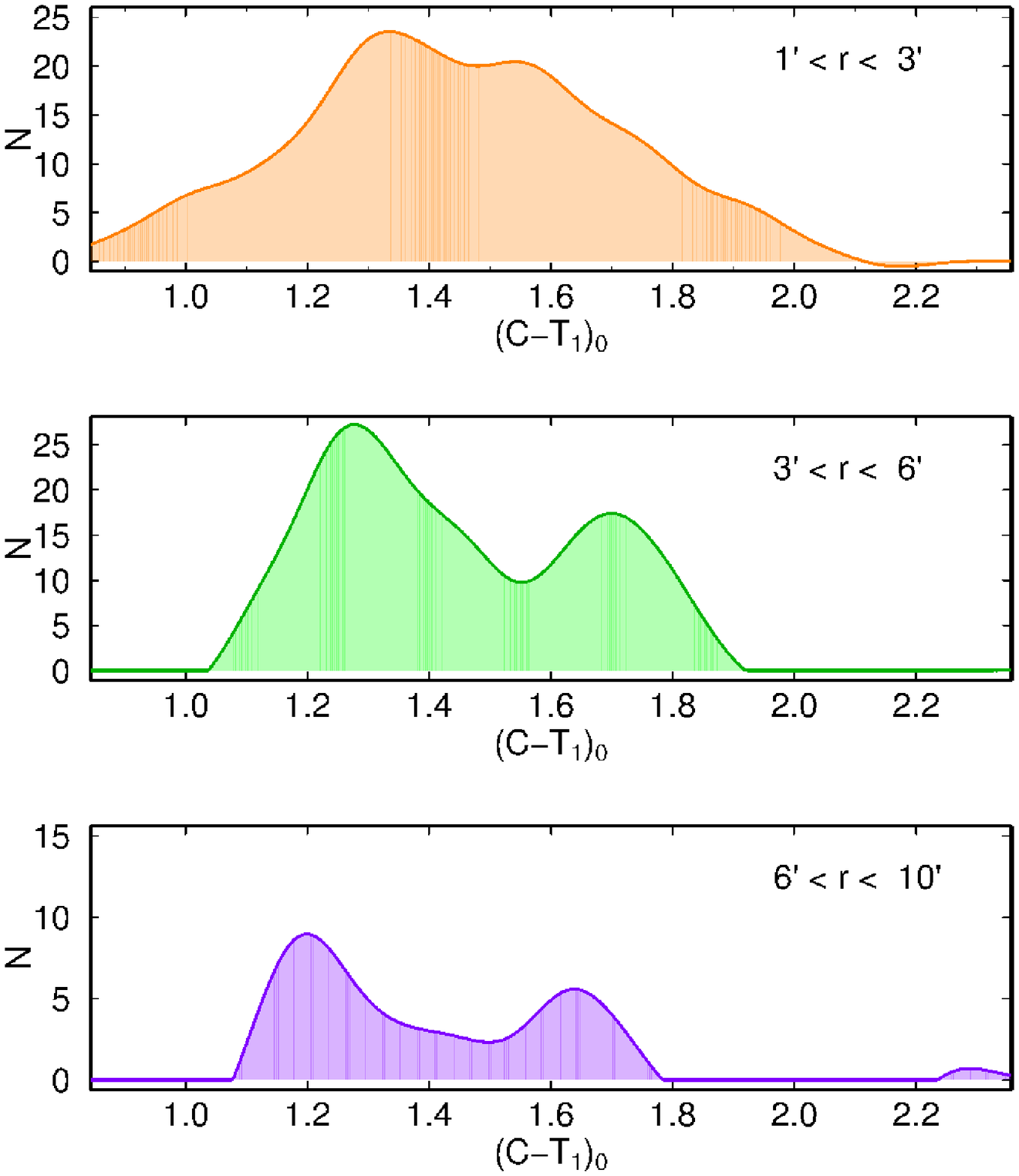}\\
\caption{Smoothed colour distribution for GC candidates around NGC\,3258
(left panel) and NGC\,3268 (right panel), split in three radial distance 
regimes. Please note the different vertical scales.}
\label{dcol}
\end{figure*}

As a consequence, the blue and red GC populations for NGC\,3258 result $6600\pm800$ and 
$1400\pm200$ members, respectively. For NGC\,3268 we obtained a population of $5200\pm700$
blue GCs and $3000\pm450$ reds. That implies that GCSs in both galaxies have similar 
richness, but their compositions differ considerably. 
The obtained populations are significantly larger than previously determined ones 
\citep{dir03b,bas08}, as expected from the larger extensions derived for both GCS in this 
paper. \citet{bas08} integrated the radial profiles up to 10\,arcmin, and obtained
populations of 6000 and 4750 GCs for NGC\,3258 and NGC\,3268, respectively. If we
integrate our radial profiles up to the same radial limit, it results in 
$\approx 6300$ and $\approx 6900$ GCs, respectively. Hence, our results are in 
agreement with \citet{bas08} ones for NGC\,3258, and the larger population is due to
the new determination of the GCS extension. On the other hand, for NGC\,3268 we obtained
a larger population in the inner $10'$.

From $V$ total magnitudes derived by the Carnegie-Irvine Galaxy Survey \citep{ho11},
foreground extintion corrections from \citet{sch11} and distance moduli from 
\citet{tul13}, absolute magnitudes for NGC\,3258 and NGC\,3268 are
$M_V=-21.5\pm0.18$ and $M_V=-21.6\pm0.15$,respectively. Hence, the specific frequencies 
\citep{har81} for NGC\,3258 and NGC\,3268 result $S_N=20.1\pm4$ 
and $S_N=18.8\pm4$, respectively. The $S_N$ values are among the largest for galaxies of 
similar luminosity, and point in both cases to evolutionary histories with a large number
of merging episodes. The fraction of red-to-blue GCs are
$f_{red}=0.18\pm0.04$ for NGC\,3258 and $f_{red}=0.38\pm0.07$ for NGC\,3268,
which is in agreement with \citet{bas08}.

\subsection{Colour distributions}
\label{colsec}

The left panels of Figure\,\ref{dcol} show the colour distributions 
for NGC\,3258 GCs, in three ranges of projected galactocentric distance. In order
to avoid contamination from the NGC\,3268 GCS, for the outer range we have
considered only GC candidates with $129\degr < PA < 309\degr$. Bimodality seems to be 
present in the two inner frames, which is expected from the extension of the 
radial profile for red GCs in Section\,\ref{radsec}. It is strinking how the blue 
peak moves towards bluer colours when the projected galactocentric distance increases. 
In order to quantify this effect, we statistically subtracted the background
contribution from the sample of GC candidates in each radial regime. Then,
we applied the algorithm Gaussian Mixture Modeling ({\sc GMM}, \citealt{mur10}) to
the clean samples, and repeated the procedure 25 times. The mean colour
for the blue GC peak in the three ranges result $1.27\pm0.01$, $1.21\pm0.01$ and 
$1.12\pm0.02$, respectively, showing that the colour change is significant.

A similar behaviour is also found in the colour distribution of NGC\,3268 
GCs (right panels of Fig\,\ref{dcol}). In this case, for the outer colour 
range we plotted GC candidates with $-51\degr < PA < 129\degr$. There are signs
of bimodality in all the frames, pointing to a wide extended subpopulation
of red GCs. From {\sc GMM}, the mean colour for the blue GC peaks are 
$1.37\pm0.02$, $1.28\pm0.02$ and $1.22\pm0.03$, respectively.

Similar colour gradients for blue GCs have been found in other giant 
ellipticals, like NGC\,1399 \citep{bas06a} or NGC\,4594 \citep{harg14}. 

A radial gradient in the colour of the blue peak can also be obtained
for both galaxies from FORS1 data, when we split the GC candidates between
those with galactocentric distances larger than $150''$ and lower than 
this limit. However, the small FOV and the limited metallicity dependence
of $(V,I)$ photometry in comparison with $(C,T_1)$ one retrict our analysis.

\subsection{Photometric metallicities}

Multicolour-metallicity relations \citep[e.g.][]{for13}, have proven to 
be useful for determining photometric metallicities for a large sample 
of GCs, with a relative low cost in observational hours.

Although an increasing number of GCs present multiple stellar populations 
\citep[e.g.][]{mil08,joo13},  we are considering globulars integrated properties 
and the age spreads between these populations are significantly smaller than the 
total ages of the GCs. Thus, single stellar populations (SSP) still constitute
a good approximation 
to GCs. Recent studies have shown that the ages of GCs indicate that they 
mostly formed at larger redshifts than $z\approx 2$ 
\citep[e.g.][and references therein]{vbe13,forb15}.
We aim at obtaining photometric metallicities 
for a subsample of NGC\,3258 GCs, based on the long-base set of photometric  
measurements available, ranging from $B$ to $z'$ bands. We assumed ages of 10, 11.2,  
and 12.6\,Gyr and metallicities in the range of $0.01 < {\rm Z/Z_{\odot}} < 0.75$. 
We used the synthetic magnitudes for SSP obtained from 
{\sc CMD\,2.8}\footnote{http://stev.oapd.inaf.it/cmd} \citep{bre12} with a 
lognormal initial mass function.

\begin{figure}
\includegraphics[width=80mm]{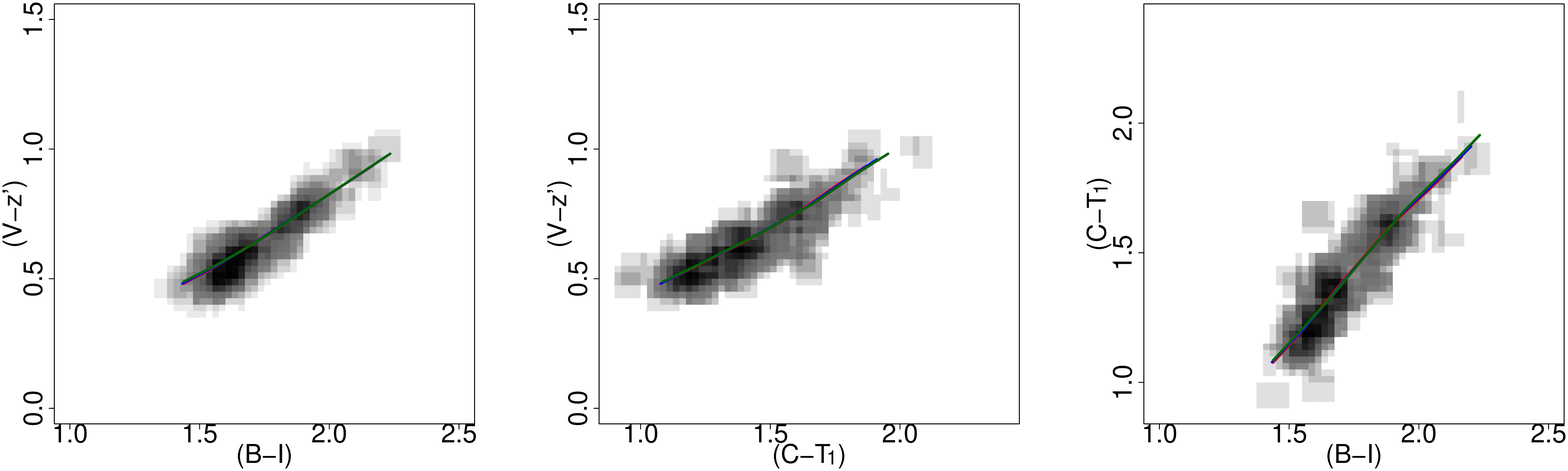}\\
\includegraphics[width=80mm]{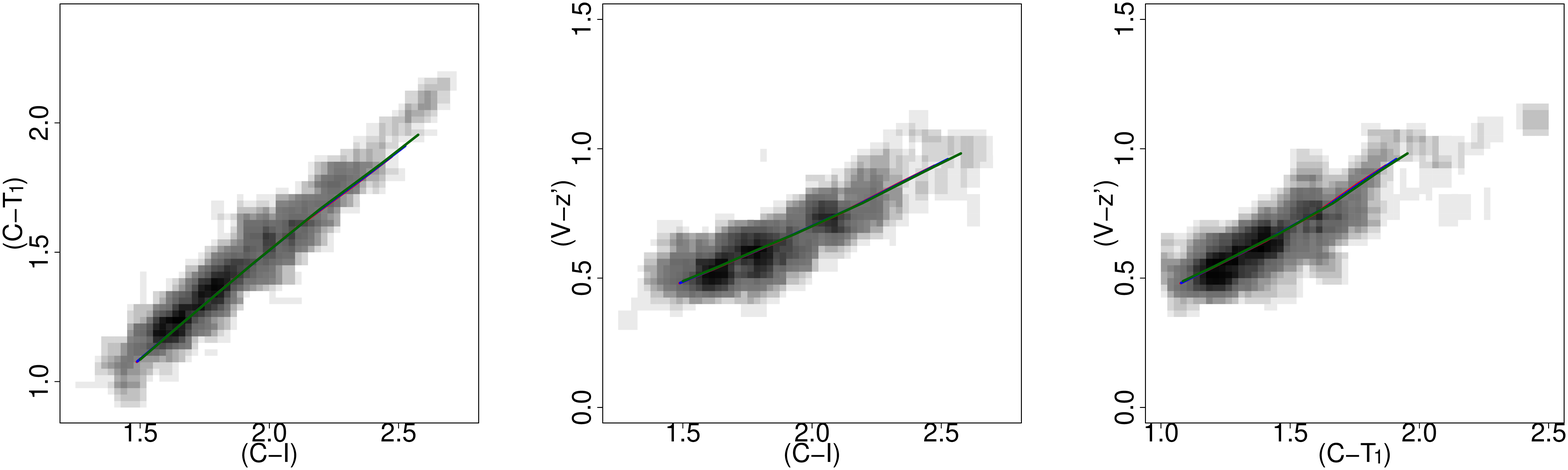}\\
\includegraphics[width=80mm]{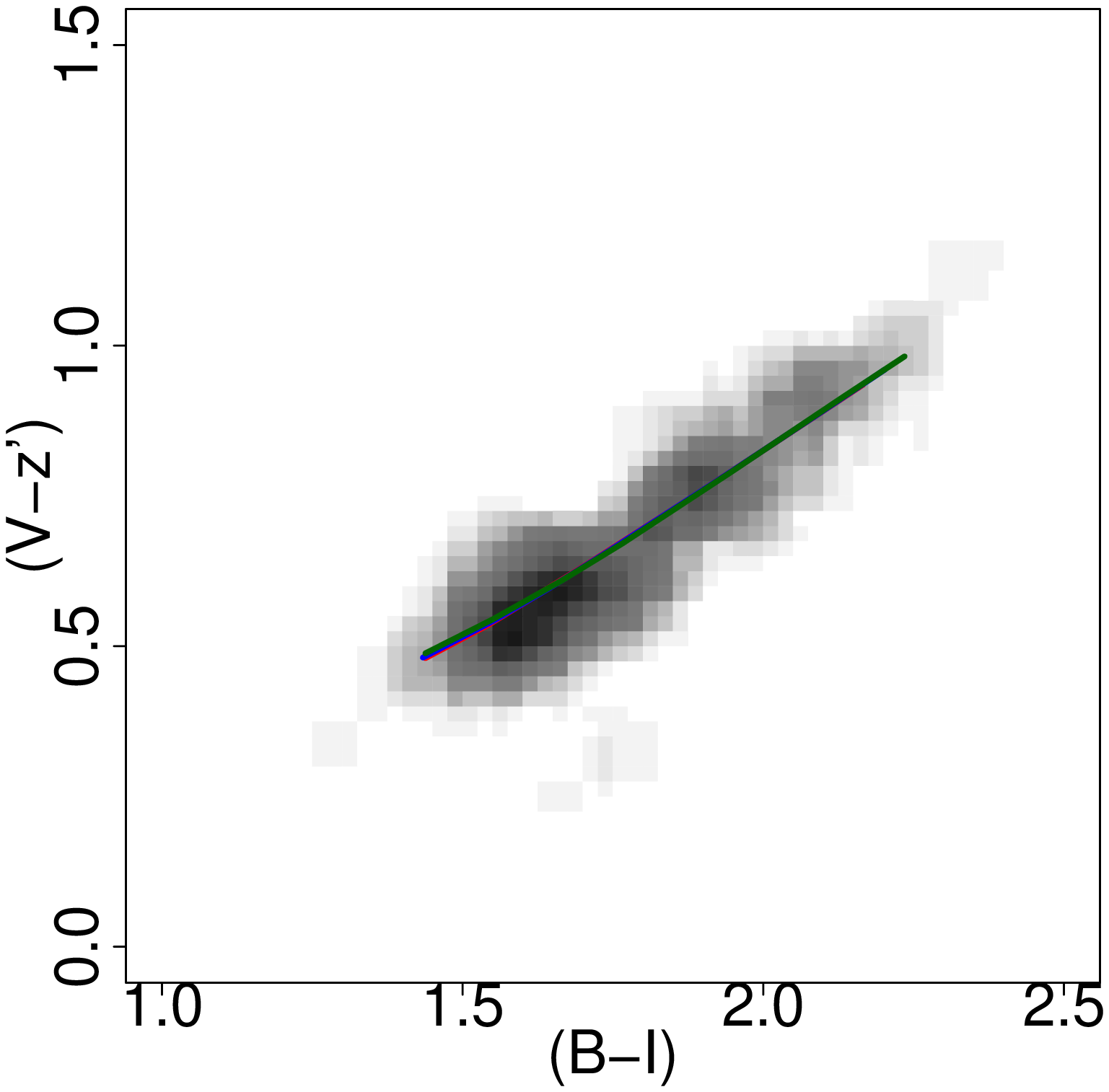}\\
\caption{Smoothed colour-colour relations defining the colour-space for groups 
A (first row), B (second row) and C (third row), depending on the available 
photometric bands (see text). The solid curves represent 
the SSP previously indicated in the text.}
\label{dcc}
\end{figure}

We selected objects brighter than $I_0= 23$, in order to achieve an acceptable 
signal-to-noise ratio. This limit implies uncertainties lower than 0.1\,mag
in the $C$ filter, 0.05\,mag in the $T_1$ filter and 0.03\,mag for the other
photometries. We split our sample in three groups, according to the set of
photometrical measurements available for each object. Objects in group A present
available photometry in $B$,$C$,$V$,$T_1$,$I$ and $z'$ filters, those in group
B lack $B$ magnitudes, and the set of magnitudes for GC candidates 
in group C corresponds to bands $B$,$V$,$I$ and $z'$. The number of GC 
candidates in the three groups are 145, 91, and 85, respectively.

For groups A and B we defined a three-dimensional space with linearly independent
colours, while a two-dimensional space was established for group C. Figure\,\ref{dcc}
shows smoothed colour-colour relations from the colour-space for each group.
The solid curves depict the SSPs from \citet{bre12} for 10 (red), 11.2 (blue), and 
12.6\,Gyr (green), respectively, presenting almost identical
positions in all the diagrams. 

Hence, ${\rm Z/Z_{\odot}}$ metallicities for each object correspond to the point in the SSP 
for which the distance to the object in the colour-space is a minimum. In order to 
obtain metallicity uncertainties, we assumed the colour uncertainties as 
dispersions of Gaussian distributions, and applied Monte-Carlo to obtain
simulated colours. For each GC candidate, we repeated the process 100 times, 
and then we calculated deviations from the resulting metallicities. 

\begin{figure*}
\includegraphics[width=80mm]{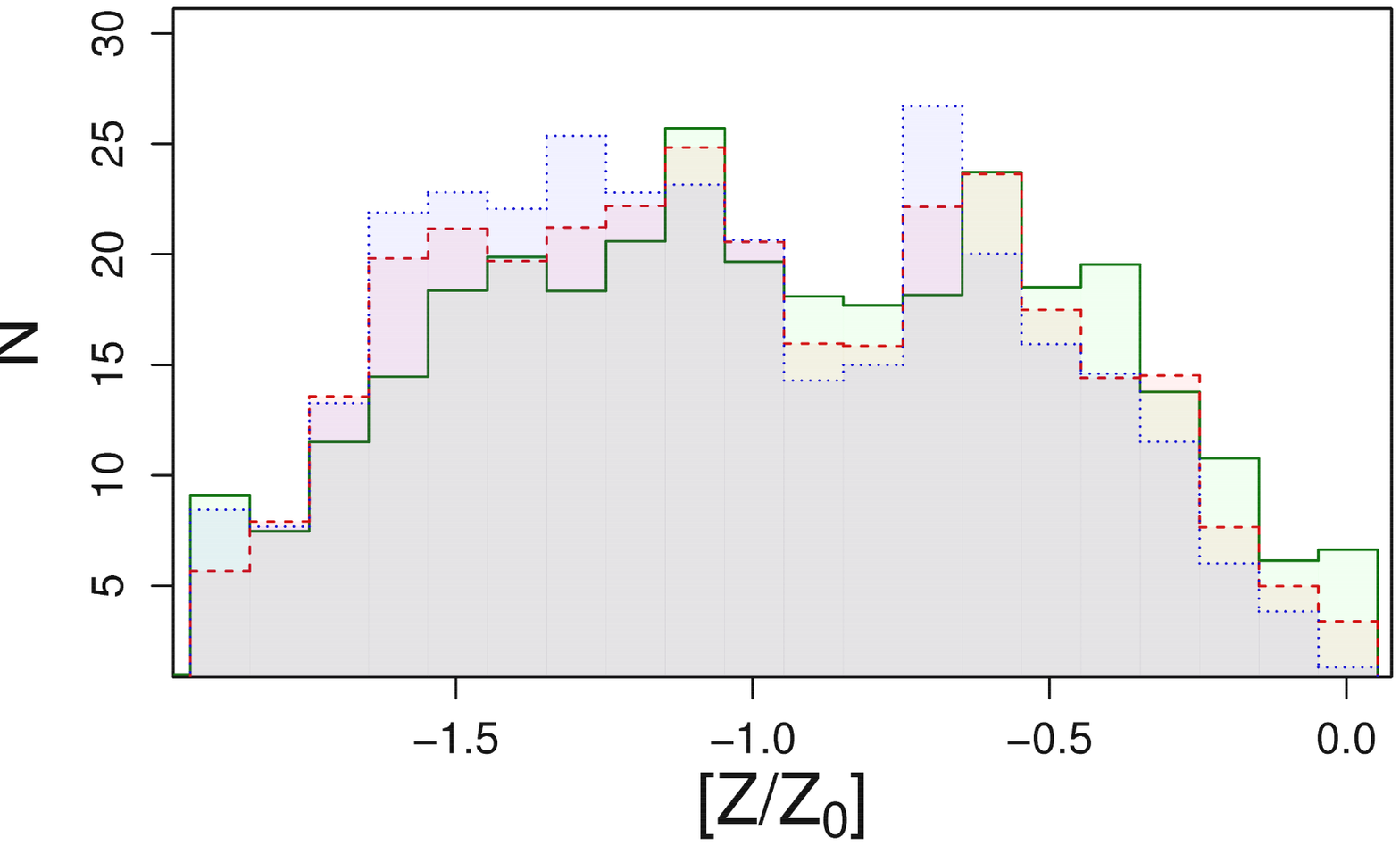}
\includegraphics[width=80mm]{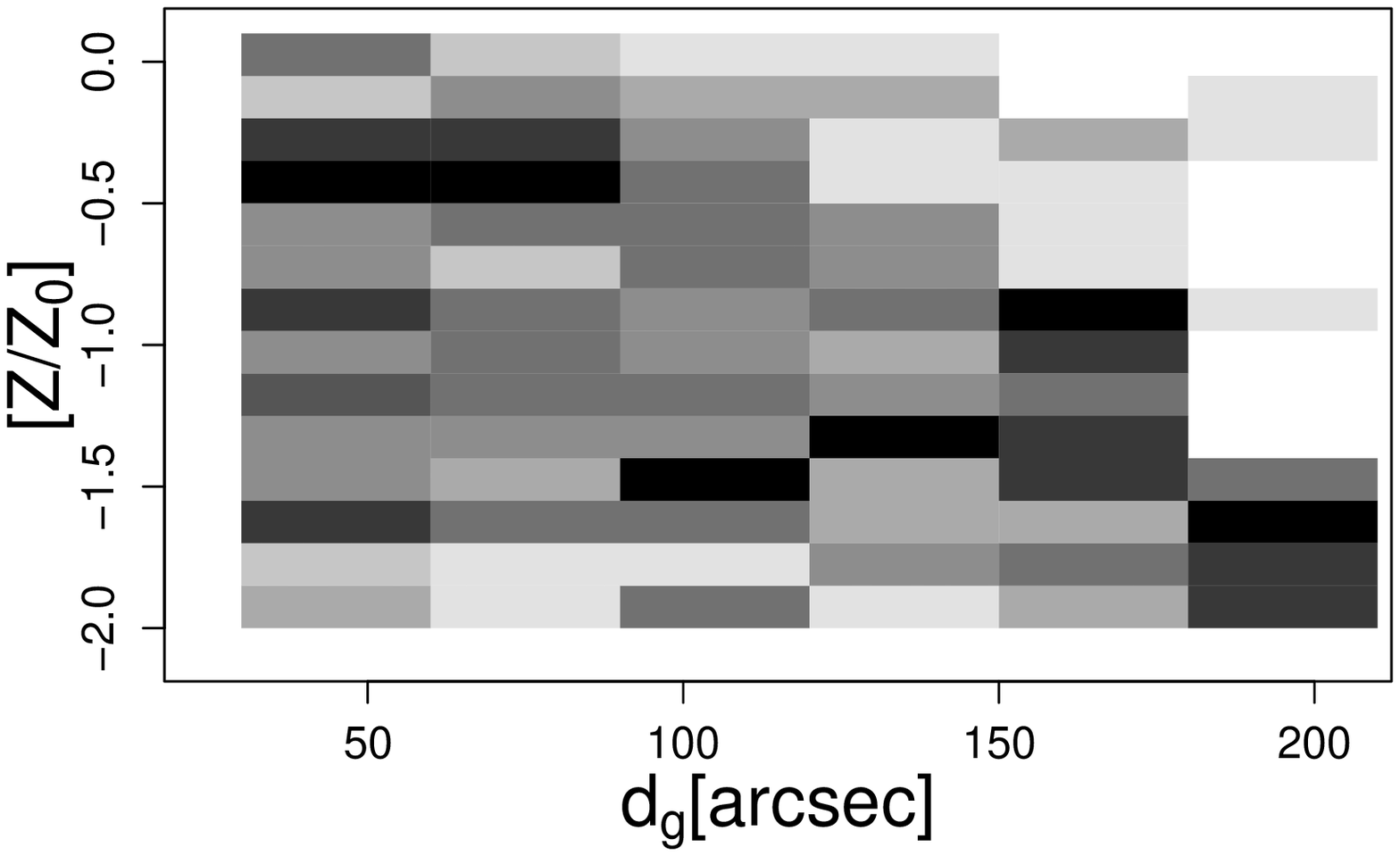}\\
\caption{{\bf Left panel:} metallicity distribution for GC candidates assuming 
SSPs with ages 10\,Gyr (solid histogram), 11.2\,Gyr (dashed histogram), and
12.6\,Gyr (dotted histogram). {\bf Right panel:} Metallicity distribution as a
function of projected distance to the centre of NGC\,3258, assuming for the GC
candidates an age of 10\,Gyr. The grey scale has been normalized for each radial bin.}
\label{dmet}
\end{figure*}

The left panel of Figure\,\ref{dmet} shows the metallicity distributions, 
assuming ages of 10\,Gyr (solid histogram), 11.2\,Gyr (dashed histogram) and 
12.6\,Gyr (dotted histogram). In the three cases, the distributions seem to 
deviate from a single Gaussian. Thus, we run {\sc GMM} to each sample, 
obtaining the results listed in Table\,\ref{gmmet}. We obtained negative 
kurtosis values and $DD$ parameters larger than 2, pointing to a bimodal 
distribution \citep{mur10}. The Gaussian peaks are in agreement within 
the uncertainties in the three cases, slightly moving towards lower 
metallicities as the age increases.

The right panel presents the metallicity distribution as a function of the 
projected distance to NGC\,3258 centre, assuming GC candidates as SSP with 
age 10\,Gyr. The grey-scale has been normalized for each radial range, with 
black representing the metallicity bins with the largest counts. As expected,
metal-rich GCs dominate close to the galaxy centre, but the mode of the 
distribution moves towards lower metallicities for larger radii.

\begin{table*}   
\begin{center}   
\caption{{\sc GMM} analysis of the metallicity distributions of GC candidates, 
assuming different ages.}    
\label{gmmet}   
\begin{tabular}{@{}ccccccc@{}}   
\hline   
\multicolumn{1}{@{}c@{}}{SSP\,[Gyr]}&\multicolumn{1}{c@{}}{${\rm [Z/Z_{\odot}]_1}$}&
\multicolumn{1}{c@{}}{${\rm [Z/Z_{\odot}]_2}$}&\multicolumn{1}{c@{}}{$\sigma_1$}&
\multicolumn{1}{c@{}}{$\sigma_2$}&\multicolumn{1}{@{}c@{}}{$DD$}&\multicolumn{1}{c@{}}{$K$}\\
\hline
$10.0$&$-1.28\pm0.11$&$-0.49\pm0.11$&$0.33\pm0.05$&$0.25\pm0.05$&$2.66\pm0.23$&$-0.97$\\
$11.2$&$-1.34\pm0.09$&$-0.56\pm0.10$&$0.30\pm0.04$&$0.26\pm0.05$&$2.80\pm0.24$&$-0.99$\\
$12.6$&$-1.36\pm0.07$&$-0.59\pm0.09$&$0.29\pm0.03$&$0.24\pm0.04$&$2.90\pm0.26$&$-0.99$\\
\hline
\end{tabular}    
\end{center}    
\end{table*}

\section{Discussion}

\subsection{Implications of the colour gradient}
\label{grad}
In the current scenario for GC formation, the blue GC population ({\it bona 
fide} metal-poor) in early-type galaxies formed during the violent 
star-formation events that occurred when first proto-galaxies begin to merge, 
or were captured from satellite galaxies. Afterwards, the metal-rich 
ones formed in merging episodes occurred in a limited number, but involving 
more massive, and more evolved, galaxies \citep{mur10,kru14,li14}. 

The accretion of dwarf galaxies could also be important to increase a GCS 
population, particularly for the metal-poor GCs 
\citep[][and references therein]{sch10,ric13,ton13}. 
Assuming this scenario, the accreted GCs could present a spatial distribution
more spread, compared with GCs formed {\it in-situ}. This would be in agreement
with the larger velocity dispersions that blue GCs present in nearby gEs 
\citep[e.g.][]{sha98,cot03,sch10,sch12,puz14}. 
Hence, the correlation 
between the peak of blue GCs in the colour distribution and galactocentric 
distance found in NGC\,3258, NGC\,3268
and NGC\,1399 \citep{bas06a} could be explained by the correlation
between the colour and metallicity for blue GCs and the galaxy masses
\citep[e.g.][]{pen06,str04a}. 

This scenario of two phases in galaxy
formation was also proposed by \citet{for11} to explain similar colour
gradients in NGC\,1407 and other giant ellipticals with large GCS like
NGC\,1399 \citep{bas06a} or M87 \citet{har09b}.

As a consequence, the radial gradient in the colour of the blue GCs peak points 
to a large accretion history of satellite galaxies for both gEs.

\subsection{Clues for the evolutionary histories in the GCS composition}
\label{clue}
Both gEs present similar halo masses \citep{ped97,nak00}, 
luminosities \citep{ho11} and $(C-T_1)$ colours 
\citep[being NGC\,3268 0.1\,mag redder][]{dir03b}. Hence their stellar masses 
should not differ significantly. From literature galaxy stellar and halo masses 
might be correlated with the size of the GCS \citep{har13,hud14}, and even 
with the fraction of red GCs \citep[$f_{red}$][]{har15}.
The higher $f_{red}$ in NGC\,3268 might indicate that it has followed  
the fiducial ellipticals' evolutionary history. 
Otherwise, the smaller $f_{red}$ in NGC\,3258 may suggest that major mergers
ocurred between galaxies with lower 
gas mass available for starburst episodes, even possibly including 
dry mergers. A mixture of wet and dry mergers could be in agreement with 
the slightly bluer colour profile of NGC\,3258 \citep{dir03b}. 
The accretion of a large number of satellite galaxies, whose GCS are
mainly formed by blue GCs \citep[e.g.][]{pen08,geo10} could also be responsible
by the smaller $f_{red}$.

\subsection{The central flattening in GCs radial profiles}
\label{flat}
 Our results for both gEs point that 
the flattening in the GC radial profiles is not explained just by the 
completeness drop, and physical processes like GCs erosion \citep{cap09}
should be invoked.
\citet{kru15} found that the same environmental conditions involved in 
GC formation favour their tidal disruption at early stages, unless its 
progenitor suffered a merger which redistributed the GCs into the 
galaxy halo. In this context, subsequent major mergers not only result 
in a larger formation of red GCs, but also improve their 
survivance ratio. Hence the larger $f_{red}$ 
and radial extension of the red GC subpopulation of NGC\,3268 with respect
to NGC\,3258 might be explained by different merger histories, as indicated
in Section\,\ref{clue}.

\section{Conclusions} 

On the basis of multi-band $B$, $C$, $V$, $R$, $I$, and $z′$ photometry, 
obtained with images taken with four different telescopes (CTIO, VLT, 
GEMINI, and archive HST), we carry out a new study of the GCSs of 
NGC\,3258 and NGC\,3268. 
Our results complement previous ones, with the main advantages 
of working with: a larger FOV, a filter combination never used before 
for these galaxies, and a more precise calculation of the completeness 
corrections. 

We summarize our conclusions in the following.
\begin{itemize}
\item Previous studies have shown that the projected total GC distributions 
are both elongated. We now confirm that blue GCs are responsible for such 
distributions, being clearly elongated in the same direction as a line 
joining both galaxies. Even an overdensity of blue GCs close to NGC\,3268 
can be detected, towards NGC\,3258. That is in agreement with the 
corresponding azimuthal GC distributions, where NGC\,3268 blue GCs show a 
peak in the direction towards NGC\,3258. Tidal effects between both galaxies 
and their blue GC subpopulations might be present, as the elongations agree 
with the elongation of the galaxies' bodies.    

\item The extension of the blue and red GC subpopulations for NGC\,3258 are 
quite different, 170\,kpc and 60\,kpc for blues and reds, respectively. However, 
for NGC\,3268 they are much more alike, 140\,kpc and 120\,kpc for blues and reds, 
respectively. The red GC subpopulation is much more concentrated in NGC\,3258 
than in NGC\,3268. 

\item Thanks to the improvement in the completeness corrections, the inner 
radial GC profiles, for blue and red subpopulations and for both galaxies, 
can be perfectly fitted with a modified Hubble law. Thus, the flattened 
inner GC profiles are pointing to a `true' effect of GC erosion in both cases.  

\item Thanks to improved GCLFs and better fits of the radial GC distributions 
for both galaxies, we estimate new GC populations. For NGC\,3258, we obtain
$6600\pm800$ blue GCs and $1400\pm200$ red GCs, and for NGC\,3268 $5200\pm
700$ blue GCs and $3000\pm450$ reds. The total GC populations are similar, 
of about 8000 members, i.e. larger than previous determinations. However, 
the fraction of red GCs is higher in NGC 3268 which suggests that, although 
both galaxies belong to the same environment, they should have experienced 
a different merging history.  

\item In the GC colour distributions, we detect a clear gradient in the 
blue GC subpopulations, in the sense that the blue peak gets bluer when 
the galactocentric radius increases. This colour gradient is present in 
NGC\,3258 as well as in NGC\,3268, and can be understood as a consequence 
of an important accretion of satellite proto/dwarf galaxies during the 
early stages of galaxy formation.

\item By means of muti-colour relations, we estimate photometric metallicities
for a subsample of GCs and fit SSPs. We recover clear bimodal metallicity 
distributions for ages between 10 and 12.6\,Gyr.
\end{itemize}

Similarities in the GCSs, i.e., richness, radial extensions and  
presence of radial gradients in the colour of blue GCs point in both cases
to a large number of accretion and merging episodes. Despite of this,
differences suggest unalike evolutionary histories of the host
galaxies, probably due to different number of dry and wet mergers.

\section*{Acknowledgments}
JPC acknowledges Tom Richtler and Francisco Azpilicueta for useful comments 
that improved the article.   

This research was funded with grants from Consejo Nacional de Investigaciones   
Cient\'{\i}ficas y T\'ecnicas de la Rep\'ublica Argentina (PIP 112-201101-00393), 
Agencia Nacional de Promoci\'on Cient\'{\i}fica y Tecnol\'ogica (PICT-2013-0317), 
and Universidad Nacional de La Plata (UNLP 11-G124), Argentina. 

\bibliographystyle{mnras}
\bibliography{biblio}

\label{lastpage}
\end{document}